\documentclass{article}
\usepackage[utf8]{inputenc}
\usepackage{authblk}
\usepackage[dvips]{graphicx}
\graphicspath{{noiseimages/}}
\usepackage{xcolor}
\usepackage[T2A]{fontenc}
\usepackage{tikz}
\usepackage{cite}
\usepackage{multirow}
\usepackage{pgfplots}
\usepackage{mathrsfs}
\usepackage{graphicx}
\usepackage{subcaption}

\pgfplotsset{compat=1.7}
\usetikzlibrary{intersections, pgfplots.fillbetween}
\usetikzlibrary{decorations.pathmorphing}
\usetikzlibrary{arrows.meta}
\usepackage[mode=buildnew]{standalone}
\usepackage[toc,page]{appendix}
\usepackage{amsmath}
\usepackage{physics}
\usepackage{amssymb}
\usepackage{float}
\usepackage{comment}
\usepackage{a4wide}

\usepackage{hyperref}
\definecolor{urlcolor}{HTML}{990000}
\definecolor{linkcolor}{HTML}{005F5F}
\hypersetup{pdfstartview=FitH,  linkcolor=linkcolor,urlcolor=urlcolor, colorlinks=true,citecolor=blue}
\usepackage[english]{babel}
\usepackage{hyperref}
\usepackage[T2A]{fontenc}
\usepackage[utf8]{inputenc}
\usepackage{bm}

\title{Superconducting Proximity Effect in an SSH-Superconductor Junction}

\author[1]{I. A. Belkovich\footnote{\tt i.belkovich@lebedev.ru}}
\author[1]{A.~A.~Radkevich\footnote{\tt radkevichaa@lebedev.ru}}

\affil[1]{\itshape P. N. Lebedev Physical Institute, Moscow 119991, Russia}

\begin{document}

\maketitle

\begin{abstract}
    A model of microscopic interaction between a superconductor and a one-dimensional topological insulator, an SSH chain, is considered. Using the functional integration method, the effective action of the interaction between a superconductor and a topological insulator is obtained. We obtain corrections to the quasiparticle excitation spectrum of the SSH chain due to tunneling in various limits and discuss the influence of phase fluctuations. We find that for bulk superconductors, the states of the chain are stable for energies lying inside the superconducting gap while in lower-dimensional superconductors phase fluctuations yield finite temperature-dependent lifetimes even inside the gap. We also discuss whether these results can be reproduced within a simple phenomenological approach.
\end{abstract}

\maketitle

\section{Introduction}
Topological materials \cite{Kane, Hasan, Xiao, Bernevig, Frank} are materials in which a nontrivial topology of electronic states leads to excitations (usually edge excitations) whose properties are protected from the influence of certain classes of perturbations. Bringing them into contact with a superconductor is one of the possible ways to obtain the so-called topological superconductors that combine topological security and superconductivity in their properties \cite{Jason, Beenakker, Zero, qubit, Spin}. The potential scope of their application is extremely wide and covers quantum computing, spintronics, magnetic field detection, and many others, and such systems are in the center of attention of modern theoretical and experimental research. While the existence of protected edge states is interesting from the point of view of their use in quantum computing, the transport and equilibrium properties of structures based on them carry new physics and are also at the center of research activities.

One of the possible uses of topological materials is the realization of topological superconductivity inside such a material by means of the superconducting proximity effect \cite{kitaev} which introduces superconducting correlations in the topological material. Despite significant progress made in this area, most theoretical research relies on two simplified approaches, each of which has significant limitations. The first widely used -- phenomenological -- approach involves directly adding the superconducting order parameter to the Hamiltonian of a topological insulator (TI) \cite{Chen, Driss, Tamura, Liang}. Although this method results in models whose properties (including symmetries and topology) can be easily analyzed, they usually lack sufficient microscopic foundation as, strictly speaking, the superconducting order parameter can only be defined in systems with attractive electron-electron interaction. Moreover, introducing a phenomenological order parameter completely ignores both space and temporal non-locality of the resulting effective theories which may affect the boundary states of the studied systems either by modifying the boundary conditions or by introducing dissipation.

The second widely used approach is based on numerical analysis of the exact electron Green's function \cite{DasSarma2010}. While this approach can be successfully used to study the influence of the induced superconductivity on the edge states of a TI, it lacks the transparency of effective models  which would allow to extend the analysis beyond the single-electron picture and account for collective phenomena.

In \cite{Giulia}, a model of dissipation of the edge states of a one-dimensional TI is discussed. The main approach is the Lindblad equation, in which the superoperator of interaction of the insulator with the environment is constructed from considerations of preservation (violation) of chiral symmetry. The authors conclude that in both situations dissipation leads to destabilization of the edge states. 

In this paper, we aim to study the proximity effect between a topological insulator and a superconductor employing an analytically tangible model and explore the consequences of the non-locality of the effective interaction induced within the TI. We pay special attention to the possible appearance of dissipation in the system both due to the tunneling and collective modes and provide an effective theory suitable to capture the influence of the superconductor on both bulk and bound states of a model TI.

\section{Set up of the problem}
In the present work, we consider a microscopic model of interaction between a massive superconductor and the simplest model of a one-dimensional topological insulator, which will be described by the so-called Su-Schrieffer-Heeger model \cite{Su}. It is a tight-binding model with hoppings between nearest neighbors, and the hoppings are alternating. Such a model exhibits the main properties of a topological insulator, namely, the presence of a forbidden zone in the bulk spectrum; the existence of a zero-energy boundary state under a certain restriction on the hopping ratio. In particular, such a model describes the electronic subsystem of polyacetylene molecules. 

The superconductor occupies half the space. The insulator lies on the surface of the superconductor. The interaction is provided by electron tunneling between the subsystems mentioned above at the point of contact. The entire system is at zero temperature. Our goal is to study the effect of inducing superconducting correlations in an insulator and how these correlations affect the edge state of the insulator.

Let us start by presenting the Hamiltonians of each of the subsystems. Su-Schrieffer-Heeger model is a one-dimensional tight-binding model with hoppings between the nearest neighbors which alternate between $t_{1}$ and $t_{2}$. For simplicity, we will consider the hoppings to be real. Within this model, a unit cell contains two atoms. The second quantized Hamiltonian of the $N$ unit cells model is written as

\begin{equation}
\label{H_SSH}
    \hat{H}_{\text{SSH}} = - \sum\limits_{m = 0}^{N-1} \sum\limits_{\sigma = \uparrow, \downarrow} \left[t_{1} \hat{a}^{\dagger}_{m, \sigma}\hat{b}_{m, \sigma} + t_{2} \hat{b}^{\dagger}_{m,\sigma} \hat{a}_{m+1, \sigma} + \text{h.c.}\right].
\end{equation}
Here $\hat{a}_{m,\sigma}, \hat{b}_{m,\sigma}$ ($\hat{a}_{m,\sigma}^{\dagger},\hat{b}_{m,\sigma}^{\dagger}$) -- fermionic annihilation (creation) operators on the first and second atom in the unit cell, respectively. We assume that the binding energy at each site is the same and set it to zero. Here we assume that the energy of excitations in the chain is measured from the chemical potential. The distance between the atoms is assumed to be equal to $a$, while the lattice period is equal to $2 a$.

Let us briefly discuss the basic properties of such a topological insulator. Firstly, the excitation spectrum of a closed chain has two branches with the law of dispersion.

\begin{equation}
\label{excitation}
    \varepsilon^{\pm}(k) = \pm \sqrt{t_{1}^2 + t_{2}^2 + 2 t_{1} t_{2} \cos{(2 k a)}},
\end{equation}
where $k$ -- momentum along the chain, taking values in the first Brillouin zone: $k \in [-\pi/2a, \pi/2a]$. Note that each branch is doubly degenerate. Secondly, one of the important properties is the presence of so-called edge states at certain hopping amplitudes. Consider a situation where the chain is semi-infinite, i.e. in (\ref{H_SSH}) $N = \infty$, and rewrite (\ref{H_SSH}) in the form of one-particle Hamiltonian

\begin{equation}
    H_{\text{SSH}} = - \sum\limits_{m = 0}^{\infty} \sum\limits_{\sigma = \uparrow, \downarrow} \left[t_{1}|m, \sigma, 1\rangle\langle m,\sigma, 2| + t_{2} |m , \sigma, 2\rangle \langle m+1, \sigma, 1| + \text{h.c.}\right]. 
\end{equation}
Here $|m, \sigma, \alpha\rangle$ -- the state in which the electron is located in the unit cell with the number $m$ on the atom $\alpha$ with spin projection $\sigma$. Moreover, $\alpha = \{1,2\}$ parameterizes the first and second atoms in the unit cell, respectively. It turns out that for such a Hamiltonian, one can find an eigenstate with zero energy that decays deep into the chain, provided $|t_{1}| < |t_{2}|$. It has the form

\begin{equation}
\label{edge}
    |\psi, \sigma\rangle = \sum\limits_{m = 0}^{\infty} \left(- \frac{t_{1}}{t_{2}}\right)^{m}\psi_{0} |m, \sigma, 1\rangle, 
\end{equation}
where $\psi_{0}$ -- the value of the wave function at the boundary of the chain, which is determined from the normalization condition.

As for the superconductor, it also needs to be written in terms of the tight-binding model, since in the future we will include the tunneling interaction with the insulator. The distance between the atoms of a superconductor is $a$, as in an insulator. In addition, we consider an s-pairing superconductor in the mean field approximation. Such a superconductor is described by the Bogolyubov Hamiltonian, which has the form

\begin{equation}
\label{H_SC}
    \begin{gathered}
        \hat{H}_{\text{SC}} = \sum\limits_{\bm{n},\bm{n}'} \sum\limits_{\sigma = \uparrow, \downarrow} \xi_{\bm{n},\bm{n'}}\hat{c}^{\dagger}_{\bm{n},\sigma} \hat{c}_{\bm{n'},\sigma}  + \sum\limits_{\bm{n}} \left[\Delta \hat{c}^{\dagger}_{\bm{n}, \uparrow}\hat{c}^{\dagger}_{\bm{n},\downarrow} + \bar{\Delta} \hat{c}_{\bm{n},\downarrow}\hat{c}_{\bm{n},\uparrow}\right].
    \end{gathered}
\end{equation}
Here $\xi_{\bm{n},\bm{n'}}$ is the kinetic contribution in the tight--binding model in the coordinate representation, calculated from the chemical potential of the metal. At the same time, $\mu$ is the chemical potential in the superconductor. Also for simplicity, the binding energy at each site is set to zero. The vector $\bm{n} = (n_{x}, n_{y}, n_{z})$ has integer components parametуrizing the position of the sites. The superconductor will occupy a half-space that is covered by coordinates with $n_{x}, n_{y} \in \mathbb{Z}$ and $n_{z} \in \{-\infty, 0\}$.  

\begin{figure}[h]
    \centering
    \includegraphics[width=0.8\textwidth]{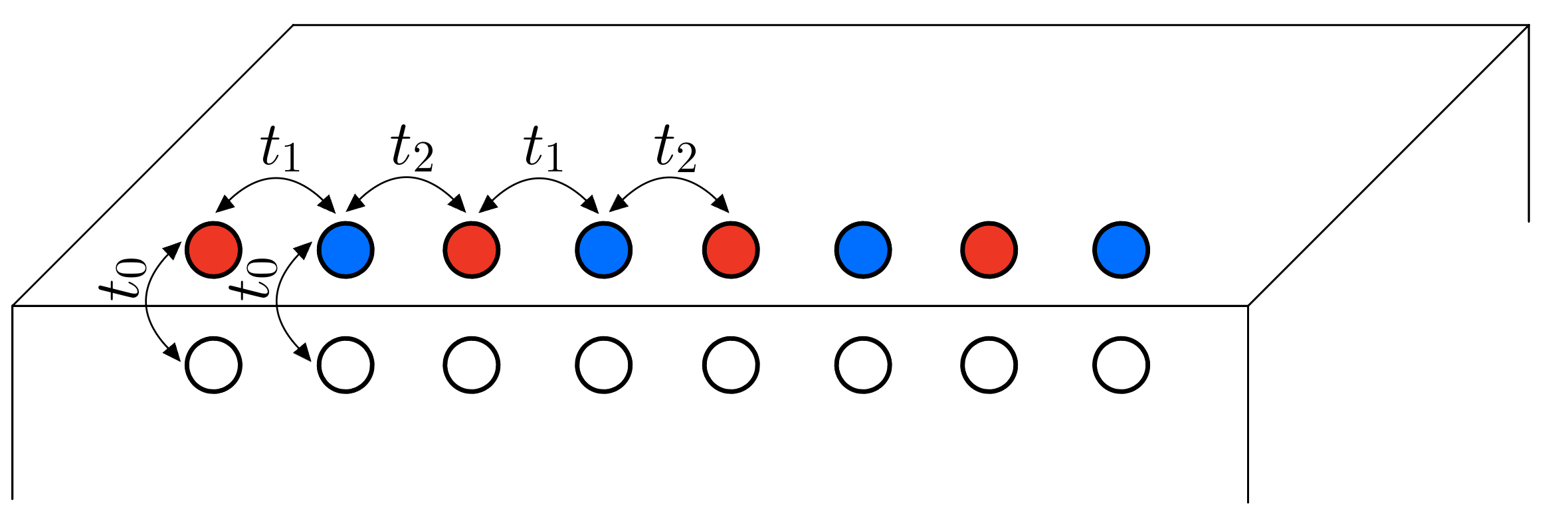}
    \caption{The considered model of a junction between a TI and a superconductor}
    \label{fig:TIandSC}
\end{figure}

Now we can proceed to the construction of a microscopic interaction of a superconductor
and an insulator. As mentioned above, the insulator lies on the surface of the superconductor, and in such a way that each atom of the insulator is located above the corresponding atom of the superconductor (see \ref{fig:TIandSC}). This possibility is provided by our choice of the same distance between atoms for both subsystems. We will consider the simplest situation when tunneling occurs only at the point of contact. The Hamiltonian describing such tunneling has the form

\begin{equation}
\label{H_{Tun}}
    \hat{H}_{\text{Tun}} = t_{0} \sum\limits_{m = 0}^{N-1}\sum\limits_{\sigma = \uparrow, \downarrow}\left[\hat{a}_{m, \sigma}^{\dagger}\hat{c}_{2m,0,0, \sigma} + \hat{b}^{\dagger}_{m, \sigma}\hat{c}_{2m+1,0,0, \sigma} + \text{h.c.}\right].
\end{equation}
Here $t_{0}$ -- real tunneling hopping, which we will assume is small in comparison with some characteristic value of energy.
Thus, the complete Hamiltonian will contain the above contributions

\begin{equation}
\label{H_total}
    \hat{H} = \hat{H}_{\text{SSH}} + \hat{H}_{\text{SC}} + \hat{H}_{\text{Tun}}.
\end{equation}

\section{Elimination of superconductor's degrees of freedom}
In the previous section, we wrote out the Hamiltonian that describes our problem. In this section, we construct an effective description of the interaction between a superconductor and a topological insulator. To achieve this goal, we use the representation of the partition function as a Grassman functional integral over coherent states. We pay attention to the fact that the Hamiltonian is quadratic with respect to fermionic creation and annihilation operators. This circumstance allows us to exactly integrate over the degrees of freedom of a superconductor and obtain an effective theory containing only the degrees of freedom of a topological insulator. From this point on, we will also assume that there is a certain binding energy, the same at each node of the chain. This energy effectively describes the difference in binding energies between the atoms of the superconductor and the atoms of the TI.

First, we need to construct a Matsubara action that corresponds to the Hamiltonian (\ref{H_total}). Let us write out what the actions look like for each of the Hamiltonians. First, we write down the action for the Hamiltonian (\ref{H_SSH}) of the SSH model with binding energy

\begin{equation}
\label{S_SSH}
    \begin{gathered}
        S_{\text{SSH}} = \int\limits_{0}^{\beta} d \tau \sum\limits_{m = 0}^{N-1}\sum\limits_{\sigma = \uparrow,\downarrow}[ \bar{\varphi}_{m, \sigma}\partial_{\tau} \varphi_{m,\sigma} +  \bar{\chi}_{m, \sigma} \partial_{\tau} \chi_{m,\sigma} - t_{1} \bar{\varphi}_{m, \sigma}\chi_{m,\sigma} -  t_{2} \bar{\chi}_{m,\sigma}\varphi_{m+1,\sigma} - \\ - t_{1} \bar{\chi}_{m,\sigma}\varphi_{m,\sigma} - t_{2} \bar{\varphi}_{m+1, \sigma}\chi_{m,\sigma} + \varepsilon_{g}\bar{\varphi}_{m,\sigma}\varphi_{m,\sigma} + \varepsilon_{g}\bar{\chi}_{m,\sigma}\chi_{m,\sigma}].
    \end{gathered}
\end{equation}
Here $\varphi_{m,\sigma}, \chi_{m,\sigma}$ -- coherent fields corresponding to the first and second atoms in the unit cell, respectively. These fields are defined as the eigenvalues of the annihilation operators, i.e. $\hat{a}_{m,\sigma} |\varphi_{m,\sigma}\rangle = \varphi_{m,\sigma} | \varphi_{m,\sigma}\rangle$ and $\hat{b}_{m,\sigma}|\chi_{m,\sigma}\rangle = \chi_{m,\sigma} | \chi_{m,\sigma} \rangle$. For the future, it turns out to be convenient to rewrite the action (\ref{S_SSH}) in terms of Nambu-Gorkov spinors which we define as

\begin{equation}
    \Phi_{m} = 
    \begin{pmatrix}
        \varphi_{m, \uparrow} \\
        \chi_{m, \uparrow} \\
        \bar{\varphi}_{m, \downarrow} \\
        \bar{\chi}_{m, \downarrow}. 
    \end{pmatrix}, \hspace{0.5cm}
    \Phi_{m}^{\dagger} = 
    \begin{pmatrix}
        \bar{\varphi}_{m,\uparrow} & \bar{\chi}_{m,\uparrow} & \varphi_{m,\downarrow} & \chi_{m,\downarrow}
    \end{pmatrix}.
\end{equation}
These spinors belong to the tensor product $\mathbf{N}\otimes \mathbf{S}$ of the Nambu and Sublattice spaces. Then

\begin{equation}
\label{S_coord}
    S_{\text{SSH}} = \int\limits_{0}^{\beta} d \tau \sum\limits_{m,m'} \Phi_{m}^{\dagger} \left[ \delta_{m,m'} \partial_{\tau} + \hat{H}^{\text{SSH}}_{m,m'}\right]\Phi_{m'},
\end{equation}
where the matrix element $ \delta_{m,m'} \partial_{\tau} + \hat{H}^{SSH}_{m,m'}$ is the inverse Matsubara Green's function of the SSH model in the coordinate representation. Let's write down the part related to the Hamiltonian 

\begin{equation*}
\label{H_coord}
    \hat{H}_{m,m'}^{\text{SSH}} = \small
    \begin{pmatrix}
        \varepsilon_{g}\delta_{m,m'} & -\delta_{m,m'}t_{1} - \delta_{m,m'+1}t_{2} & 0 & 0 \\
        - \delta_{m,m'}t_{1} - \delta_{m', m+1} t_{2} & \varepsilon_{g}\delta_{m,m'} & 0 & 0 \\
        0 & 0 & -\varepsilon_{g}\delta_{m,m'} & \delta_{m,m'}t_{1}+\delta_{m,m'+1}t_{2} \\
        0 & 0 & +\delta_{m,m'}t_{1}+\delta_{m',m+1}t_{2} & -\varepsilon_{g}\delta_{m,m'} 
    \end{pmatrix}.
\end{equation*}

Similarly, for the Hamiltonian of a superconductor (\ref{H_SC})

\begin{equation}
    \begin{gathered}
        S_{\text{SC}} = \int\limits_{0}^{\beta} d \tau \sum\limits_{\bm{n},\bm{n}'} \sum\limits_{\sigma = \uparrow, \downarrow}[\bar{\psi}_{\bm{n},\sigma}\partial_{\tau} \psi_{\bm{n},\sigma} + \xi_{\bm{n},\bm{n}'}\bar{\psi}_{\bm{n},\sigma}\psi_{\bm{n},\sigma}] + \int\limits_{0}^{\beta} d \tau \sum\limits_{\bm{n}} [\Delta \bar{\psi}_{\bm{n},\uparrow}\bar{\psi}_{\bm{n},\downarrow} + \bar{\Delta} \psi_{\bm{n},\downarrow}\psi_{\bm{n},\uparrow}]
    \end{gathered}
\end{equation}
Where $\psi_{\bm{n},\sigma}$ - coherent field, which is defined as $\hat{c}_{\bm{n},\sigma} | \psi_{\bm{n},\sigma}\rangle = \psi_{\bm{n},\sigma}|\psi_{\bm{n},\sigma}\rangle$. We rewrite this action in terms of Nambu-Gorkov spinors

\begin{equation}
    \Psi_{\bm{n}} = 
    \begin{pmatrix}
        \psi_{\bm{n},\uparrow} \\
        \bar{\psi}_{\bm{n},\downarrow} 
    \end{pmatrix},
    \hspace{0.5cm}
    \Psi_{\bm{n}}^{\dagger} = 
    \begin{pmatrix}
        \bar{\psi}_{\bm{n},\uparrow} & \psi_{\bm{n}, \downarrow}
    \end{pmatrix}.
\end{equation}
Then

\begin{equation}
\label{S_{SC}}
    S_{\text{SC}} = \int d t \sum\limits_{\bm{n},\bm{n}'} \Psi_{\bm{n}}^{\dagger}\left[ \delta_{\bm{n},\bm{n}'}\partial_{\tau} + \hat{H}^{SC}_{\bm{n},\bm{n}'}\right] \Psi_{\bm{n}'}.
\end{equation}
As before $ \delta_{\bm{n},\bm{n}'}\partial_{\tau} + \hat{H}^{SC}_{\bm{n},\bm{n}'}$ is the inverse Matsubara Green's function of superconductor that knows the boundary conditions (\ref{j_{z}}).As in the previous case, we will write out the matrix elements of the Hamiltonian

\begin{equation}
\label{H_SC}
    \hat{H}_{\bm{n},\bm{n}'}^{\text{SC}} = 
    \begin{pmatrix}
        \xi_{\bm{n},\bm{n}'} & \delta_{\bm{n},\bm{n}'}\Delta \\
        \delta_{\bm{n},\bm{n}'}\bar{\Delta} & - \xi_{\bm{n},\bm{n}'}
    \end{pmatrix}
\end{equation}

Finally, we can write out the action responsible for tunneling between the two subsystems which corresponds to the Hamiltonian (\ref{H_{Tun}}). We will immediately write it out in finite form in terms of Nambu-Gorkov spinors for SSH and superconductor

\begin{equation}
    S_{\text{Tun}} = \int\limits_{0}^{\beta} d\tau \sum\limits_{\bm{n}, m} \left[\Psi^{\dagger}_{\bm{n}} \hat{T}_{\bm{n}, m}\Phi_{m} + \Phi_{m}^{\dagger}\hat{T}_{m,\bm{n}}^{\dagger}\Psi_{\bm{n}}\right],
\end{equation}
where $\hat{T}_{\bm{n},m}$ -- tunneling operator which corresponding to the transitions between the site of the superconductor with the number $\bm{n}$ and site of the SSH chain with the number $m$. It has the form

\begin{equation}
\label{Tun}
    \hat{T}_{\bm{n},m} = t_{0} \delta_{n_{y},0}\delta_{n_{z},0}
    \begin{pmatrix}
         \delta_{n_{x},2m} &  \delta_{n_{x},2m+1} & 0 & 0 \\
        0 & 0 & -\delta_{n_{x},2m} & -\delta_{n_{x},2m+1}
    \end{pmatrix}.
\end{equation}
Thus, we have constructed an action for each Hamiltonian, and now we can write down the full action

\begin{equation}
    S = S_{\text{SSH}} + S_{\text{SC}} + S_{\text{Tun}}.
\end{equation}

Now we can represent the partition function as a Grassmann functional integral over coherent states \cite{Altland}

\begin{equation}
    \mathcal{Z} = \int \mathcal{D} [\bar{\eta}, \eta] e^{-S[\bar{\eta},\eta]},
\end{equation}
where $\eta = \{\psi_{\sigma}, \varphi_{\sigma}, \chi_{\sigma}\}$ -- a set of all fermion fields that describe the system. It should be noted that these Grassmann functions obey anti-periodic boundary conditions $\eta(0) = -\eta(\beta)$. Next, we will assume that partition function integrates over Nambu-Gorkov spinors. Now we can proceed to calculate the effective action. To do this, in the representation of the partition function in terms of Nambu-Gorkov spinors, we integrate the degrees of freedom of the superconductor

\begin{equation}
    \mathcal{Z} = \int \mathcal{D}[\Psi, \Psi^{\dagger}, \Phi, \Phi^{\dagger}] e^{-S} = \det\left[\delta_{\bm{n},\bm{n}'} \partial_{\tau} +  \hat{H}^{\text{SC}}_{\bm{n},\bm{n}'}\right] \int \mathcal{D}[\Phi, \Phi^{\dagger}] e^{- S_{\text{eff}}}.
\end{equation}
Thus we obtained the effective action, which contains only degrees of freedom of the SSH chain. It is given

\begin{equation}
\label{S_{eff}}
    S_{\text{eff}} = S_{\text{SSH}} + \int\limits_{0}^{\beta} d \tau \int\limits_{0}^{\beta} d \tau' \sum\limits_{\bm{n}, \bm{n}'} \sum\limits_{m, m'} \Phi_{m}^{\dagger}(\tau) \hat{T}_{m, \bm{n}}^{\dagger} \hat{G}^{\text{M}}_{\bm{n},\bm{n}'}(\tau-\tau') \hat{T}_{\bm{n}', m'} \Phi_{m'}(\tau'), 
\end{equation}
where $G^{\text{M}}$ is the Green's function of the superconductor satisfying
\begin{equation}
    \left(\partial_{\tau} +  \hat{H}^{\text{SC}}_{\bm{n},\bm{n}'}\right)\hat{G}^{\text{M}}_{\bm{n}',\bm{n}''}(\tau-\tau') = \hat{1}\delta_{\bm{n},\bm{n}''}\delta(\tau-\tau')\label{G^M_def}
\end{equation}

The last term of Eq.(\ref{S_{eff}}) corresponds to the effective interaction between the two subsystems. As mentioned above, the Green's function $\hat{G}^{M}$ of the superconductor satisfies the boundary conditions 
\begin{equation}
\label{j_{z}}
    j_{z} = 0, \hspace{0.5cm} \text{on the surface}.
\end{equation}
In this case, it can be represented through a sum of two free Green's functions $\hat{\tilde{G}}$ of a superconductor occupying the entire space as
\begin{equation}
\label{boundary_Green}
    \hat{G}^{\text{M}}_{\bm{n},\bm{n}'} = \left[\hat{\tilde{G}}^{\text{M}}_{n_{x}-n_{x}', n_{y}-n_{y}', n_{z}-n_{z}'} + \hat{\tilde{G}}^{\text{M}}_{n_{x}-n_{x}', n_{y}-n_{y}', -n_{z}+n_{z}'}\right]/2
\end{equation}
Notice that free Green function $\hat{G}^{\text{M}}_{\bm{n}-\bm{n}'}$ has a matrix structure in the  Nambu space

\begin{equation}
\label{Green_matrix}
    \hat{G}^{\text{M}}_{\bm{n}-\bm{n}'} =
    \begin{pmatrix}
        G^{\text{M}\uparrow \uparrow}_{\bm{n}-\bm{n}'} & G^{\text{M}\uparrow \downarrow}_{\bm{n}-\bm{n}'} \\
        G^{\text{M}\downarrow \uparrow}_{\bm{n}-\bm{n}'} & G^{\text{M}\downarrow \downarrow}_{\bm{n}-\bm{n}'}
    \end{pmatrix}.
\end{equation}
Using the rations (\ref{boundary_Green}) and (\ref{Green_matrix}) we can partially calculate the sum in Eq.(\ref{S_{eff}}) and get the final expression for an effective action.

\begin{equation}
\label{S_eff_fin}
    S_{\text{eff}} = S_{\text{SSH}} + \int\limits_{0}^{\beta} d \tau \int\limits_{0}^{\beta} d \tau' \sum\limits_{m,m'}\Phi_{m}^{\dagger}(\tau) \hat{V}_{m,m'}(\tau-\tau') \Phi_{m'}(\tau'),
\end{equation}
where the interaction operator has the form

\begin{equation}
\begin{gathered}
\label{V_{int}}
    \hat{V}_{m,m'} = \hat{T}_{m, \bm{n}}^{\dagger} \hat{G}^{\text{M}}_{\bm{n},\bm{n}'} \hat{T}_{\bm{n}', m'} = \\ = t_{0}^2
    \begin{pmatrix}
            G_{2m-2m',0,0}^{\uparrow \uparrow} & G_{2m-1-2m',0,0}^{\uparrow \uparrow} & - G_{2m-2m',0,0}^{\uparrow \downarrow} & - G_{2m-1-2m',0,0}^{\uparrow \downarrow} \\ 
            G_{2m+1-2m',0,0}^{\uparrow \uparrow} & G_{2m-2m',0,0}^{\uparrow \uparrow} & -G_{2m+1-2m',0,0}^{\uparrow \downarrow} & - G_{2m-2m',0,0}^{\uparrow \downarrow} \\ 
            -G_{2m-2m',0,0}^{\downarrow \uparrow} & - G_{2m-1-2m',0,0}^{\downarrow \uparrow} &  G_{2m-2m',0,0}^{\downarrow \downarrow} &  G_{2m-1-2m',0,0}^{\downarrow \downarrow} \\
            - G_{2m+1-2m',0,0}^{\downarrow \uparrow} & - G_{2m-2m',0,0}^{\downarrow \uparrow} &  G_{2m+1-2m',0,0}^{\downarrow \downarrow} & G_{2m-2m',0,0}^{\downarrow \downarrow}
        \end{pmatrix}
\end{gathered}
\end{equation}
Here we have omitted the index $M$ and dependence on $\tau-\tau'$ for brevity. As you can see the interaction (\ref{V_{int}}) allows the electrons on the chain to tunnel not only to nearby neighboring sites, but also to longer distances. At the same time, due to superconducting correlations or anomalous Green's functions, it becomes possible to hop between sites with a spin flip, which leads to mixing of spin states.

Let us rewrite Eq.(\ref{S_eff_fin}) in frequency representation

\begin{equation}
\label{S_F}
\begin{gathered}
    S_{\text{eff}} = T \sum\limits_{n} \sum\limits_{m,m'} \Phi^{\dagger}_{m}(i\omega_{n})\left[-\delta_{m,m'}i\omega_{n} + \hat{H}_{m,m'}^{\text{SSH}} + \hat{V}_{m,m'}(i\omega_{n}) \right]\Phi_{m'}(i\omega_{n}),
\end{gathered}
\end{equation}
where $\hat{V}_{m,m'}(i\omega_{n})$ -- Fourier image of interaction operator.

We have obtained an expression for an effective action in terms of the Matsubara representation. As aware, using the analytical continuation procedure, one can obtain a retarded Green's function from the Matsubara. Since our problem is quadratic in fermionic fields, this procedure consists in simply replacing Matsubara frequencies $\omega_{n}$ with real frequencies $\omega$ according to the rule: $i\omega_{n} \rightarrow \omega + i 0$. Performing such a transition, we obtain real-frequency representation of an action (\ref{S_F})

\begin{equation}
\label{S_K}
    S_{\text{eff}} = \int\limits_{-\infty}^{+\infty} \frac{d \omega}{2 \pi} \sum\limits_{m,m'}\Phi_{m}^{\dagger}(\omega)\left[\delta_{m,m'}\omega - \hat{H}_{m,m'}^{\text{SSH}} -\hat{V}_{m,m'}(\omega)\right]\Phi_{m'}(\omega). 
\end{equation}
The action (\ref{S_K}) will be used in further calculations of the energy spectrum of a topological insulator, as well as in studying the effects of superconducting correlations on the edge state.

\section{Bulk states}
In the previous section, we obtained an expression (\ref{S_eff_fin}) for effective action, which contains the action of the chain and the action of the effective interaction of the chain and the superconductor. This action contains only the degrees of freedom of the insulator. In this section, we will discuss how the presence of effective interaction affects the properties of excitations in the bulk of a topological insulator.

Since we will now be interested in excitations in the bulk, we may assume the chain to be infinite, i.e. consider $m,m' \in \mathbb{Z}$ in Eq.(\ref{S_K}). This assumption provides an opportunity to perform a Fourier transform in position space for (\ref{S_K}). Since we will be exploring the spectrum of the system, we will proceed to the frequency representation. After all we obtain 

\begin{equation}
\label{S_k}
    S_{\text{eff}} = \int\limits_{-\infty}^{+\infty} \frac{d \omega}{2 \pi} \int\limits_{-\frac{\pi}{2a}}^{\frac{\pi}{2a}} \frac{a d k}{\pi} \Phi^{\dagger}_{\omega,k} \left[\omega - \hat{H}^{\text{SSH}}_{k} - \hat{V}_{\omega,k}\right] \Phi_{,\omega,k}.
\end{equation}
Here $k$ -- a momentum along the insulator chain. We will also write out separately the matrices for the free Hamiltonian

\begin{equation}
\label{H_k}
    \hat{H}^{\text{SSH}}_{k} =
    \begin{pmatrix}
        \varepsilon_{g} & -t_{1} - t_{2} e^{-2ika} & 0 & 0 \\
        -t_{1} - t_{2} e^{2ika} & \varepsilon_{g} & 0 & 0 \\
        0 & 0 & -\varepsilon_{g} &  t_{1} + t_{2} e^{-2i k a} \\
        0 & 0 &  t_{1} + t_{2} e^{2i k a} & -\varepsilon_{g} 
    \end{pmatrix},
\end{equation}
and for the interaction term 

\begin{equation}
\label{V_eff}
\begin{gathered}
    \hat{V}_{\omega,k} = t_0^2 \int \frac{a^2 d^2 \bm{k}_{\perp}}{(2\pi)^2}  \times \\ \times \small{
    \begin{pmatrix}
        G_{k}^{\uparrow\uparrow}+G^{\uparrow\uparrow}_{k+\frac{\pi}{a}} & [G_{k}^{\uparrow\uparrow} -G^{\uparrow\uparrow}_{k+\frac{\pi}{a}}]e^{-ika} & -G_{k}^{\uparrow\downarrow}-G^{\uparrow\downarrow}_{k+\frac{\pi}{a}} & [-G^{\uparrow\downarrow}_{k}+G^{\uparrow\downarrow}_{k+\frac{\pi}{a}}]e^{-ika} \\
        [G_{k}^{\uparrow\uparrow} -G^{\uparrow\uparrow}_{k+\frac{\pi}{a}}]e^{ika} & G_{k}^{\uparrow\uparrow}+G^{\uparrow\uparrow}_{k+\frac{\pi}{a}} & [-G^{\uparrow\downarrow}_{k}+G^{\uparrow\downarrow}_{k+\frac{\pi}{a}}]e^{ika} & -G_{k}^{\uparrow\downarrow}-G^{\uparrow\downarrow}_{k+\frac{\pi}{a}} \\ -
        G_{k}^{\downarrow\uparrow}-G^{\downarrow\uparrow}_{k+\frac{\pi}{a}} & [-G^{\downarrow\uparrow}_{k}+G^{\downarrow\uparrow}_{k+\frac{\pi}{a}}]e^{-ika} & G_{k}^{\downarrow\downarrow}+G^{\downarrow\downarrow}_{k+\frac{\pi}{a}} & [G^{\downarrow\downarrow}_{k} -G^{\downarrow\downarrow}_{k+\frac{\pi}{a}}]e^{-ika} \\
        [-G^{\downarrow\uparrow}_{k}+G^{\downarrow\uparrow}_{k+\frac{\pi}{a}}]e^{ika} & -G_{k}^{\downarrow\uparrow}-G^{\downarrow\uparrow}_{k+\frac{\pi}{a}} & [G^{\downarrow\downarrow}_{k}-G^{\downarrow\downarrow}_{k+\frac{\pi}{a}}]e^{ika} & G_{k}^{\downarrow\downarrow}+G^{\downarrow\downarrow}_{k+\frac{\pi}{a}}.
    \end{pmatrix}}
\end{gathered}
\end{equation}
Here $\bm{k}_{\perp}$ -- a projection of the momentum, which is perpendicular to the chain. At the same time the integration takes place over the area with $\bm{k}_{\perp} \in [-\frac{\pi}{a}, \frac{\pi}{a}]\times[-\frac{\pi}{a}, \frac{\pi}{a}]$. Here we omitted the index $\bm{k}_{\perp}$ of the Green's function for brevity. Also, the Fourier transform for the retarded Green's function of superconductor has the form

\begin{equation}
\label{G}
    \hat{G}^{\text{R}}_{\omega,\bm{k}} = \frac{1}{(\omega + i 0)^2 - \xi_{\bm{k}}^2- |\Delta|^2}
    \begin{pmatrix}
        \omega + \xi_{\bm{k}} & \Delta \\
        \bar{\Delta} & \omega - \xi_{\bm{k}}
    \end{pmatrix}
\end{equation}
where $\xi_{\bm{k}} = \varepsilon_{\bm{k}} - \mu$ -- the energy of quasiparticle excitations in a metal, calculated from the Fermi level. 

The hermitian conjugate $\hat{V}^\dag$ can be found by using Eq.(\ref{V_eff}) upon substituting the retarded Green's function $\hat{G}^{\text{R}}$ with its advanced counterpart $\hat{G}^{\text{A}}$ (here we used the fact that $(\hat{G}^{\text{R}})^\dagger = \hat{G}^{\text{A}})$. Therefore, the effective interaction is hermitian as long as $\hat{G}^{\text{R}} = \hat{G}^{\text{A}}$ which is satisfied for the energies lying within the superconducting gap. Outside of the gap, $\hat{G}^{\text{R}}$ acquires a non-hermitian part given by 
\begin{equation}
\label{Imw}
    \text{Im} \hat{G}^{\text{R}}_{\bm{k}}(\omega) = - \pi \text{sign}(\omega) \delta(\omega^2 - \xi_{\bm{k}}^2 - |\Delta|^2)\begin{pmatrix}
        \omega + \xi_{\bm{k}} & \Delta \\
        \bar{\Delta} & \omega - \xi_{\bm{k}}.
    \end{pmatrix}.
\end{equation}
These observations imply that at energies inside the superconducting gap electron states are localized in the vicinity of the chain despite the tunneling inside the superconductor and that they have infinite lifetimes despite the time-non-local nature of $\hat{V}$. This happens as a result of absence of superconducting quasiparticle states at those energies. On the other hand, states with energies outside of the superconducting gap have finite lifetimes which means that electrons initially localized at the chain will eventually mix with some quasiparticle state inside a superconductor and go to infinity.

The spectrum of the system is determined by the poles of the system's Green's function and can be found by solving
\begin{equation}
    \det\left[\omega - \hat{H}^{\text{SSH}}_{k} - \hat{V}_{\omega,k}\right] = 0.
\end{equation}
Note that in the absence of an interaction operator, this equation yields an unperturbed dispersion law of the form (\ref{excitation}). We will solve this equation using perturbation theory in a small parameter $t_{0}^2$.

Before we move on to the results of calculation of corrections to the energy spectrum of states in the bulk, let us discuss the symmetry properties of the effective Hamiltonian of our system, derived from microscopic considerations, and the constraints they impose on the kind of the spectrum. First, let us note that the spectrum of a superconductor and the spectrum of the SSH model are symmetric with respect to zero energy. This is because these subsystems exhibit chiral symmetry. Indeed, for the Hamiltonian (\ref{H_k}), one can consider the operator

\begin{equation}
\label{U_SSH}
    \hat{\mathcal{U}}_{\text{SSH}} = \hat{\sigma}_{y}\otimes \hat{1} = 
    \begin{pmatrix}
        \hat{O}^{2\times2} & -i\hat{1}^{2\times2} \\
        i\hat{1}^{2\times2} & \hat{O}^{2\times2}
    \end{pmatrix},
\end{equation}
where $\hat{O}^{2\times2}$ is the zero matrix of size $2\times2$, and $\hat{1}^{2\times2}$ is the identity matrix of size $2\times2$. This operator anticommutes with the Hamiltonian (\ref{H_k}). Similarly, for the Hamiltonian (\ref{H_SC}), one can also present an operator that anticommutes with it; it has the following form

\begin{equation}
\label{U_SC}
    \hat{\mathcal{U}}_{\text{SC}} = 
    \begin{pmatrix}
        0 & -i e^{i\varphi} \\
        i e^{-i\varphi} & 0
    \end{pmatrix},
\end{equation}
where $\varphi$ - the phase of the order parameter, defined as $\Delta = |\Delta|e^{i\varphi}$. It is straightforward to verify that the total Hamiltonian (\ref{H_total}), rewritten in the basis of Nambu-Gorkov spinors $(\Phi, \Psi)^{T}$

\begin{equation}
    \hat{H} = 
    \begin{pmatrix}
        \hat{H}_{\text{SSH}} & \hat{T} \\
        \hat{T}^{\dagger} & \hat{H}_{\text{SC}}
    \end{pmatrix},
\end{equation}
exhibit chiral symmetry, constructed using operators (\ref{U_SSH}) and (\ref{U_SC})

\begin{equation}
\label{U_total}
    \hat{\mathcal{U}} = \hat{\mathcal{U}}_{\text{SSH}}\otimes \hat{1}_{\text{SC}} + \hat{1}_{\text{SSH}}\otimes \hat{\mathcal{U}}_{\text{SC}},
\end{equation} 
where $\hat{1}_{SC}$ and $\hat{1}_{SSH}$ - identity operators, acting on the space of spinors $\Psi$ and $\Phi$, respectively. The tunneling operator $\hat{T}$ is defined in Eq.(\ref{Tun}). In turn, $\mathcal{U}_{SSH}$ now has an additional dependence on the phase of the order parameter

\begin{equation}
    \hat{\mathcal{U}}_{\text{SSH}} = 
    \begin{pmatrix}
        \hat{O}^{2\times2} & -ie^{i\varphi}\hat{1}^{2\times2} \\
        i e^{-i\varphi} \hat{1}^{2\times2} & \hat{O}^{2\times2}
    \end{pmatrix}
\end{equation}
The existence of (\ref{U_total}) implies that the interacting system continues to exhibit a symmetric spectrum with respect to the zero energy value. Moreover, after integrating out the superconductor variables, this chiral symmetry does not disappear. However, the operator of this symmetry will take a somewhat more complex form. This transformed chiral symmetry operator will anticommute with the effective Hamiltonian that includes (\ref{V_eff}).

\begin{figure}[h]
    \centering
    
    \begin{subfigure}{0.45\textwidth}
        \centering
        \includegraphics[width=\textwidth]{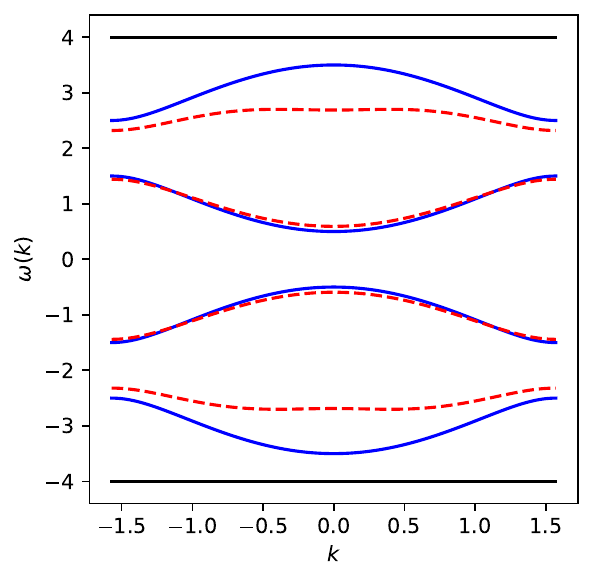}
        \caption{$\varepsilon_g \neq 0$}
        \label{fig:img1}
    \end{subfigure}
    \hfill
    \begin{subfigure}{0.45\textwidth}
        \centering
        \includegraphics[width=\textwidth]{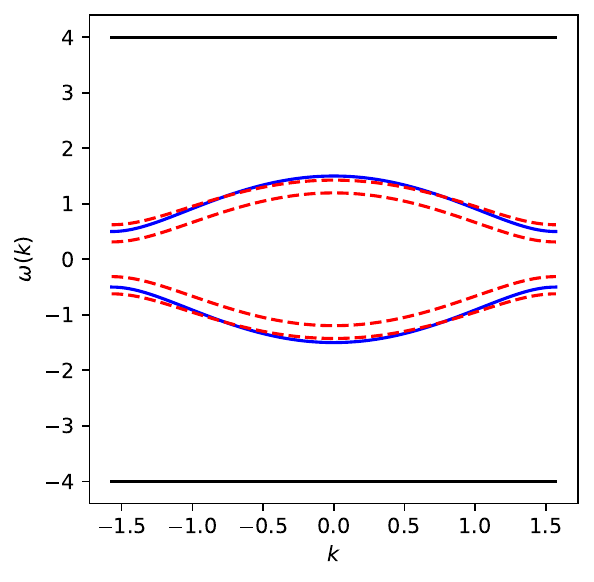}
        \caption{$\varepsilon_{g} = 0$}
        \label{fig:img2}
    \end{subfigure}
    
    \caption{Perturbative numerical calculation of the spectrum. For the superconductor we used the model spectrum of a cubic lattice $\xi_k = -2t[\cos{(k_x a)}+\cos{(k_y a)}+\cos{(k_z a)}]-\mu$ with the following parameters: (a) $\varepsilon_{g} = 2$, $t_{1} = 1$, $t_{2} = 0.5$, $ t_{0} = 0.8$, $t = 1$, $\Delta = 4$, $\mu = 0$; (b) $\varepsilon_{g} = 0$, $t_{1} = 1$, $t_{2} = 0.5$, $ t_{0} = 0.8$, $t = 1$, $\Delta = 4$, $\mu = 1$. The branches of the unperturbed spectrum are shown in blue, while the spectrum with corrections is shown in red. The black horizontal lines mark the boundaries of the superconductor gap region.}
    \label{fig:both}
\end{figure}

Now we can move on to the results of the corrections to the energy spectrum in the bulk. The unperturbed Hamiltonian (\ref{H_k}) has four branches of the energy spectrum: two of them correspond to electron states with spin $\uparrow$ with energy $\omega_{1/2} = \varepsilon_{g} \pm \omega_{0}$ and the remaining two correspond to hole states with spin $\downarrow$ with energy $\omega_{3/4} = -\varepsilon_{g} \pm \omega_{0}$, where $\omega_{0} = \sqrt{t_{1}^2+t_{2}^2+2t_{1}t_{2}\cos{(2ka)}}$ --- the law of dispersion (\ref{excitation}) without binding energy. 

At finite binding energy we use non-degenerate perturbation theory in a small parameter $t_0^2$ for each branch of the spectrum. In the leading order we obtain the following expressions for the spectrum with corrections

\begin{equation}
\label{bulk_excitation}
\begin{gathered}
    \omega_1 = \varepsilon_{g} + \omega_{0} + I(k, \omega_{0}) + I\left(k+\frac{\pi}{a}, \omega_{0}\right) \\
    \omega_2 = \varepsilon_{g} - \omega_{0} + I(k, -\omega_{0}) + I\left(k+\frac{\pi}{a}, -\omega_{0}\right),
\end{gathered}
\end{equation}
where we introduced the notation for the integral

\begin{equation}
    I(k,\pm\omega_{0}) = t_{0}^2\left[1 \mp \frac{(t_{1}+t_{2})\cos{(ka)}}{\omega_{0}}\right] \int \frac{a^2 d^2 \bm{k}_{\perp}}{(2\pi)^2} \frac{\pm\omega_{0} + \varepsilon_{g} + \xi_{k,\bm{k}_{\perp}}}{(\pm\omega_{0} + \varepsilon_{g}+i0)^2 - \xi_{k,\bm{k}_{\perp}}^2 - |\Delta|^2}. 
\end{equation}
Due to chiral symmetry, the remaining two branches of the spectrum are expressed through those written above as $\omega_{3} = -\omega_{1}$ and $\omega_{4} = - \omega_{2}$. Note that at zero binding energy, the unperturbed branches of the spectrum merge into two branches symmetrical with respect to the zero-energy value. In this case, degenerate perturbation theory should be used to calculate the corrections. You can find the procedure for calculating the corrections in both cases in Appendix A.

Figure (\ref{fig:both}) shows numerical graphs of the spectrum taking into account energy corrections when the energies lie inside the superconducting gap, i.e. $|\omega_{0}+\varepsilon_{g}| < |\Delta|$. In this energy region, the corrections are real, which is due to the Hermitian nature of the effective interaction (\ref{V_eff}) in this region. Note that the perturbation removes the degeneracy of the spectrum in the system with $\varepsilon_{g} = 0$. We will return to this observation in Section 6. It is also worth mentioning that, depending on the precise spectrum and chemical potential of the superconductor, the correction may assume any sign, while for the simple spectrum of the cubic lattice at half filling which we used for modeling it is negative.

In the opposite limit, when we are outside the gap, the corrections to the spectrum begin to contain an imaginary part $\text{Im}(\omega) \sim -t_0^2\nu_0 a^3$ (see Eq.(\ref{im}) in the Appendix), which is associated with the finite lifetime of quasiparticles through the standard uncertainty relation $\tau \sim - 1/\text{Im}(\omega)$.

\section{Edge states}
In the previous section, we calculated corrections to the excitation spectrum of the chain in the bulk. In this section, we discuss the effect of the interaction with a superconductor on the properties of the edge state of the topological insulator. As before we will be interested in the Hamiltonian (\ref{H_SSH}) in the general case, when the binding energy at each site is taken into account. That is, the Hamiltonian (\ref{H_SSH}) will be rewritten as

\begin{equation}
\begin{gathered}
\label{H_g}
    \hat{H}_{\text{SSH}} = - \sum\limits_{m = 0}^{N-1} \sum\limits_{\sigma = \uparrow, \downarrow} \left[t_{1} \hat{a}^{\dagger}_{m, \sigma}\hat{b}_{m, \sigma} + t_{2} \hat{b}^{\dagger}_{m,\sigma} \hat{a}_{m+1, \sigma} + \text{h.c.}\right] + \\ + \sum\limits_{n = 0}^{N-1}\sum\limits_{\sigma = \uparrow,\downarrow}\left[\varepsilon_{g}\hat{a}^{\dagger}_{n,\sigma}\hat{a}_{n,\sigma} + \varepsilon_{g} \hat{b}^{\dagger}_{n,\sigma} \hat{b}_{n,\sigma}\right].
\end{gathered}
\end{equation}
We assume that the bond energies of the atoms in the unit cell are the same. Note that in this case, the edge state energy is also shifted by the amount $\varepsilon_{g}$. Keeping in mind (\ref{edge}), the equation for the unperturbed edge state now reads

\begin{equation}
    \hat{H}_{\text{SSH}} |\psi,\sigma\rangle = \varepsilon_{g} |\psi, \sigma\rangle.
\end{equation}
We will be interested in the case where the energy $\varepsilon_{g}$ lies within the gap of a superconductor.

As in Section 2, we will consider a semi-infinite chain that begins with the first atom in a unit cell. To answer the question about the fate of the edge state, we will use an effective action (\ref{S_K}) that is written in the real space representation. In the frequency-coordinate representation, it will take the form

\begin{equation}
\label{S_2}
    S_{\text{eff}} = \int \frac{d \omega}{2 \pi} \sum\limits_{m, m' = 0}^{\infty} \Phi_{m}^{\dagger} \left[\omega \delta_{m,m'} - \hat{H}_{m,m'}^{\text{SSH}} - \hat{V}_{\omega,m,m'}\right]\Phi_{m'}.
\end{equation}
Here $\hat{H}_{m,m'}^{SSH}$ is the same as in (\ref{S_coord}) except for the additional contribution from the binding energy. 

We will find corrections to the ground-state energy using perturbation theory. Therefore, we need to rewrite the edge state vectors (\ref{edge}) in terms of  Nambu-Gorkov spinors for up and down spin as follows

\begin{equation}
\label{Edge}
    |\psi, \uparrow\rangle \longrightarrow \Phi_{m}^{\uparrow} =  \psi_{0} \left(-e^{-2 \kappa a}\right)^{m}
    \begin{pmatrix}
        1 \\ 0 \\ 0 \\ 0
    \end{pmatrix}, \hspace{0.5cm} 
    |\psi, \downarrow\rangle \longrightarrow \Phi_{m}^{\downarrow} =  \psi_{0} \left(-e^{-2 \kappa a}\right)^{m}
    \begin{pmatrix}
        0 \\ 0 \\ 1 \\ 0
    \end{pmatrix}.
\end{equation}
Here we introduced a notation $t_{1}/t_{2} = e^{-2 \kappa a}$, where $\kappa$ makes sense for the decay coefficient on the lattice period. $\psi_{0}$ is determined from
the normalization condition 

\begin{equation}
    \sum\limits_{m = 0}^{\infty} \Phi^{\sigma\dagger}_{m} \Phi_{m}^{\sigma'} = \delta_{\sigma, \sigma'}, \hspace{0.5cm} \Rightarrow \hspace{0.5cm} \psi_{0}^2 = 1 - e^{-4 \kappa a}.
\end{equation}

\begin{equation}
\label{V2D}
    \hat{V}_{\text{2D}} = 
    \begin{pmatrix}
        V_{\uparrow \uparrow} & V_{\uparrow \downarrow} \\
        V_{\downarrow \uparrow} & V_{\downarrow \downarrow}
    \end{pmatrix}, \hspace{0.25cm} \text{where} \hspace{0.25cm} V_{\sigma \sigma'} = \sum\limits_{m,m' = 0}^{\infty} \Phi^{\sigma \dagger}_{m} \hat{V}_{m,m'} \Phi_{m}^{\sigma'}.
\end{equation}
Let's write out this matrix elements

\begin{equation}
    V_{\sigma \sigma^\prime} = t_{0}^2\sum\limits_{m,m' = 0}^{\infty} G_{2m-2m', 0, 0}^{\sigma \sigma^\prime}(\omega) \psi_{0}^2 (-e^{-2\kappa a})^{m+m'},
\end{equation}
where $\sigma$, $\sigma^\prime$ can assume values $\uparrow,\downarrow$ or, equivalently, $+$ and $-$. By representing the superconductor's Green function in terms of Fourier transform, matrix elements can be rewritten as

\begin{equation}
    V_{\sigma \sigma^\prime} =  t_{0}^2  \sum\limits_{m,m' = 0}^{+\infty} \int\limits_{-\frac{\pi}{a}}^{\frac{\pi}{a}} \frac{a d k}{2\pi} \int \frac{a^2 d \bm{k}_{\perp}}{(2\pi)^2} G_{k, \bm{k}_{\perp}}^{\sigma \sigma^\prime} (\omega) \psi_{0}^2 (-1)^{m+m'} e^{i 2 k a (m-m') -2 \kappa a(m+m')}.
\end{equation}
Taking the sum over the site indexes $m,m'$ and substituting the value $\psi_{0}^2$, we obtain 

\begin{equation}
    V_{\sigma\sigma^\prime} = (-1)^{\sigma\sigma^\prime}t_{0}^2 \int\limits_{-\frac{\pi}{a}}^{\frac{\pi}{a}} \frac{a d k}{2\pi} \int \frac{a^2 d \bm{k}_{\perp}}{(2\pi)^2} G_{k, \bm{k}_{\perp}}^{\sigma\sigma^\prime} (\omega) \frac{\sinh{(2\kappa a)}}{\cosh{(2\kappa a)} + \cos{(2 k a)}}
\end{equation}
The kernel of the integral (\ref{S_2}) will have the following form in the basis of edge state vectors

\begin{equation}
    \omega \delta_{m,m'} - \hat{H}^{\text{SSH}}_{m,m'} - \hat{V}_{m,m'} \hspace{0.25cm} \rightarrow \hspace{0.25cm} 
    \begin{pmatrix}
        \omega - \varepsilon_{g} & 0 \\
        0 & \omega + \varepsilon_{g}
    \end{pmatrix}
    - 
    \begin{pmatrix}
        V_{\uparrow\uparrow} & V_{\uparrow\downarrow} \\
        V_{\downarrow\uparrow} & V_{\downarrow\downarrow}
    \end{pmatrix}
\end{equation}
In the leading order by $t_{0}^2$ we obtain

\begin{equation}
    \begin{gathered}
        \omega - \varepsilon_{g} - V_{\uparrow\uparrow}(\omega) = 0, \\
        \omega + \varepsilon_{g} - V_{\downarrow\downarrow}(\omega) = 0.
    \end{gathered}
\end{equation}
Solving this equation and taking into account that $V_{\uparrow\uparrow}(\omega) = - V_{\downarrow\downarrow}(-\omega)$, we obtain

\begin{equation}
\varepsilon = \pm (\varepsilon_{g} + V_{\uparrow\uparrow}(\varepsilon_{g})) = \pm \left(\varepsilon_{g} + t_{0}^2 \int \frac{a^3 d^3 \bm{k}}{(2\pi)^3} \frac{\varepsilon_{g} + \xi_{k}}{\varepsilon_{g}^2 - \xi_{k}^2 - |\Delta|^2} \frac{\sinh{(2\kappa a)}}{\cosh{(2\kappa a)} + \cos{(2 k a)}}\right).
\label{edge_state_correction_1}
\end{equation}
Again, the "minus" sign is an artifact of the particle-hole transformation incorporated in the Nambu spinor formalism. Therefore, the energy of the edge state acquires the same correction for both spin projections. The obtained expression diverges at $\varepsilon_g \rightarrow \Delta$ as $\varepsilon - \varepsilon_g \sim t_0^2\nu_0 a^3 \sqrt{\frac{\Delta}{\Delta - \varepsilon_g}}$, hence, it ceases to be applicable at $\varepsilon_g \sim (t_0^2\nu_0 a^3)^{2/3} \Delta^{1/3}$ where the perturbation becomes large and the wavefunction of the edge state changes significantly. In this case, one needs to seek for the appropriate solution for the operator in action Eq.(\ref{S_F}) numerically.

If $\varepsilon_g \rightarrow 0$, then the integral in Eq.(\ref{edge_state_correction_1}) is strongly suppressed by the approximate electron-hole symmetry in the vicinity of the Fermi energy. If, in addition, the edge state is sufficiently well localized, i.e., $\exp{-2\kappa a} \ll 1$, then the edge state energy approaches the value
\begin{equation}
    \varepsilon = \sqrt{ V_{\uparrow\downarrow}V_{\downarrow\uparrow}} = t_{0}^2 \int \frac{a^3 d^3 \bm{k}}{(2\pi)^3} \frac{|\Delta|}{\xi_{k}^2 + |\Delta|^2} \frac{\sinh{(2\kappa a)}}{\cosh{(2\kappa a)} + \cos{(2 k a)}}.
    \label{edge_state_correction_2}
\end{equation}

\section{Effect of the collective modes}
Besides quasiparticles, the superconductor also hosts a set of collective modes \cite{kulik1981}, among which the most physically impactful is the plasma mode associated with smooth variations of the phase of the order parameter. In lower-dimensional samples, i.e., a sufficiently thin wire, the spectrum of this mode is gapless \cite{mooij1985} which allows it to influence the physics of the system down to the lowest temperatures \cite{SCin1D}. In particular, it was shown that quantum fluctuations of phase yield non-zero electron density of states $\nu(E)$ inside the superconducting gap which depends on the energy $E$ as $\nu (E) \propto \nu_0\exp ((E-\Delta)/T)$, see\cite{RSZ17,RSZ19}. For our setup, this means that if the superconductor is sufficiently thin in one direction, then even the states inside the gap will acquire a finite lifetime. In order to see this, let us use Eq.(\ref{S_{eff}}) and the fact that in the leading order (taking into account only massless fluctuations) the Green's function $G^{\text{SC}}$ of a low-dimensional superconductor in the presence of a non-trivial phase configuration is given by 
\begin{equation}
    \hat{G}^{\text{SC}}(\tau,\tau',x,x') = e^{\frac{i}{2}\varphi(\tau,x)\hat{\sigma}_{z}} \hat{G}^{\text{BCS}}(\tau,\tau',x,x')e^{-\frac{i}{2}\varphi(\tau',x')\hat{\sigma}_{z}}
\end{equation}
where $G^{\text{BCS}}$ is the unperturbed BCS Green's function defined by Eq.(\ref{G^M_def}). Now, using Eqs.(\ref{V_eff},\ref{S_F}), we get
\begin{equation}
    \hat{G}[\varphi] = \left((\hat{G}^{\text{SSH}})^{-1} - \hat{T}^\dagger \hat{G}^{\text{SC}} \hat{T}\right)^{-1}
\end{equation}
valid in the presence of phase fluctuations. Now, expanding this relation in powers of $\hat{G}^{\text{SC}}$ and performing the averaging over phase fluctuations, we immediately see that the self-energy $\hat{\Sigma}$ defined as  
\begin{equation}
    \langle\hat{G}[\varphi]\rangle_{\varphi} = \left((\hat{G}^{\text{SSH}})^{-1} - \hat{\Sigma}\right)^{-1}\label{Sigma_def},
\end{equation}
in the leading order in the tunneling is given by
\begin{equation}
    \hat{\Sigma}^{(1)} = \hat{T}^\dagger\langle \hat{G}^{\text{SC}} \rangle_{\varphi}\hat{T}.
\end{equation}
In essence, we see that phase fluctuations may be accounted for by simply averaging the superconductor's Green's function over phase, thus, we can perform such a replacement in Eq.(\ref{V_eff}). Now, $V_{\text{eff}}$ is no longer a hermitian operator for $\omega$ in the superconducting gap due to the fluctuation-induced density of subgap states \cite{RSZ17,RSZ19}. Hence, the states of the chain acquire a finite lifetime $\Gamma^{-1}$ set by the imaginary part $\Gamma$ of the energy,
\begin{equation}
\label{lifetime}
    \Gamma(\omega) \sim t_0^2 a^3 \nu_0 \exp ((\omega-\Delta)/T).
\end{equation}
This phenomenon affects the boundary state of the SSH chain in the same manner it affects the bulk states. Therefore, all the states of the chain are destabilized by interaction with the collective modes of a lower-dimensional superconductor which allows them to escape into the superconductor by absorbing enough energy from thermally-excited superconducting plasmons.

A more thorough calculation along the lines of \cite{RSZ17,RSZ19} is required in order to get an accurate functional dependence of the lifetime on $\omega$ closer to the gap edge, however, this task falls beyond the scope of the present work. We note that in 3-dimensional superconductors the phase mode acquires a large gap (see \cite{kulik1981}) which further suppresses the effect and leaves the quasiparticle lifetimes virtually infinite.

\section{Naive proximity effect}
Quite often, a purely phenomenological approach is used to describe the appearance of superconducting correlations inside a material as a result of the proximity effect. Within this approach, a phenomenological order parameter $\Delta$ is introduced at each site of the topological insulator's lattice, as it would appear in the BCS theory. This approach is the most widespread (see, e.g. \cite{Driss}) because it allows one to ignore the labor-intensive calculation of induced superconducting correlations and therefore eliminates the spatial and temporal dependence of the effective interaction between the superconductor and the material under consideration. In reality, the order parameter or the BCS pairing field appears only in materials with attractive effective interaction between electrons while the superconducting correlations may penetrate the material without such an attraction.  Let us see whether a toy model with a phenomenologically introduced $\Delta$ (a "naive" model) can reproduce the results obtained within the microscopic approach. 

Consider the Hamiltonian of the SSH model (\ref{H_g}) to which we will add the contribution corresponding to the superconducting correlations

\begin{equation}
\label{naive}
\begin{gathered}
    \hat{H}_{\text{SSH}+\Delta} = \sum\limits_{n = 0}^{N-1} \sum\limits_{\sigma = \uparrow, \downarrow} \left[-t_{1} \hat{a}^{\dagger}_{n, \sigma}\hat{b}_{n, \sigma} - t_{2} \hat{b}^{\dagger}_{n,\sigma} \hat{a}_{n+1, \sigma}\right] +\sum\limits_{n = 0}^{N-1} \left[\Delta \hat{a}^{\dagger}_{n,\uparrow}\hat{a}^{\dagger}_{n,\downarrow} + \Delta \hat{b}^{\dagger}_{n,\uparrow}\hat{b}^{\dagger}_{n,\downarrow}\right] + \text{h.c.} + \\ + \sum\limits_{n = 0}^{N-1}\sum\limits_{\sigma = \uparrow,\downarrow}\left[\varepsilon_{g}\hat{a}^{\dagger}_{n,\sigma}\hat{a}_{n,\sigma} + \varepsilon_{g} \hat{b}^{\dagger}_{n,\sigma} \hat{b}_{n,\sigma}\right].
\end{gathered}
\end{equation}
Note that the $\Delta$ introduced here is not equal to the superconductor's order parameter, but rather should be set equal to $V_{\uparrow\downarrow}$. Let us consider a closed chain. Then, passing into the momentum representation of the Hamiltonian (\ref{naive}) and writing it in the basis of Nambu-Gorkov spinors we obtain

\begin{equation}
\label{naive_k}
    \hat{H}^{\text{SSH}+\Delta}_{k} = 
    \begin{pmatrix}
        \varepsilon_{g} & -t_{1} - t_{2} e^{-2ika} & \Delta & 0 \\
        -t_{1} - t_{2} e^{2ika} & \varepsilon_{g} & 0 & \Delta \\
        \bar{\Delta} & 0 & -\varepsilon_{g} &  t_{1} + t_{2} e^{-2i k a} \\
        0 & \bar{\Delta} &  t_{1} + t_{2} e^{2i k a} & -\varepsilon_{g} 
    \end{pmatrix}
\end{equation}
Where $k$ is the same as in (\ref{excitation}). Before we calculate the bulk energy spectrum of this model, we would like to draw attention to the peculiarity that this model of induced superconductivity exhibits. 
Let us note that this effective Hamiltonian also exhibits chiral symmetry, which can be realized by the operator (\ref{U_SSH}). It turns out when $\varepsilon_{g} = 0$, this model has an extra symmetry of the following form, written in the basis of Nambu-Gorkov spinors

\begin{equation}
    \hat{\Gamma} = 
    \begin{pmatrix}
        \hat{O}^{2\times2} & e^{i\varphi }\hat{\sigma}_{z} \\
        e^{-i\varphi} \hat{\sigma}_{z} & \hat{O}^{2\times2}
    \end{pmatrix}
    \label{naive_symmetry}
\end{equation}
This operator commutes with the Hamiltonian (\ref{naive_k}) when $\varepsilon_{g} = 0$, leading to degeneracy of its energy spectrum. At the same time, the effective interaction (\ref{V_eff}), derived from microscopic calculations, does not exhibit such symmetry even in the case of zero binding energy. Consequently, (\ref{V_eff}) reduces the degeneracy and splits the energy spectrum of the SSH model when coupled to a superconductor, what can be seen in the picture (\ref{fig:both}).

The excitation spectrum of (\ref{naive_k}) has four branches with the law of dispersion

\begin{equation}
\label{enaive}
\begin{gathered}
    \varepsilon_{1} = \sqrt{(\omega_{0} + \varepsilon_{g})^2+|\Delta|^2}, \hspace{1cm} \varepsilon_{2} = \sqrt{(\omega_{0} -\varepsilon_{g})^2+|\Delta|^2} \\
    \varepsilon_{3} = - \sqrt{(\omega_{0} - \varepsilon_{g})^2+|\Delta|^2}, \hspace{1cm} \varepsilon_{4} = - \sqrt{(\omega_{0}+\varepsilon_{g})^2+|\Delta|^2}
\end{gathered}
\end{equation}
Note that the expressions under the root is always positive, i.e., even in the case of normal metal ($\Delta = 0$) there will be no decay of excitations in the chain, in contrast to our model of microscopic interaction, which is taken into account.

It remains to discuss the effect of such superconducting correlation engineering on the edge state energy. To do this, we consider a semi-infinite chain, in which we add the binding energy, as in (\ref{H_g}). As before, let us write down the action of such a system in the frequency-coordinate representation

\begin{equation}
\label{S_ssh}
    S_{\text{SSH}+\Delta} = \int \frac{d \omega}{2\pi} \sum\limits_{m} \Phi_{m}^{\dagger}\left[\omega \delta_{m,m'}- \hat{H}^{\text{SSH}}_{m,m'} -\hat{V}\delta_{m,m'}\right]\Phi_{m'},
\end{equation}
where the matrix of perturbation has the form
\begin{equation}
    \hat{V} =
    \begin{pmatrix}
        0 & 0 & \Delta & 0 \\
        0 & 0 & 0 & \Delta \\
        \bar{\Delta} & 0 & 0 & 0 \\
        0 & \bar{\Delta} & 0 & 0
    \end{pmatrix}.
\end{equation}
As in Section 5, we will search for corrections to the energy of the edge states (\ref{Edge}). Let us project the perturbation operator onto the subspace of the edge state vectors

\begin{equation}
    \hat{V}_{\text{edge}} = 
    \begin{pmatrix}
        V_{\uparrow \uparrow} & V_{\uparrow \downarrow} \\
        V_{\downarrow \uparrow} & V_{\downarrow \downarrow}
    \end{pmatrix} = 
    \begin{pmatrix}
        0 & \Delta \\
        \bar{\Delta} & 0
    \end{pmatrix}.
\end{equation}
The kernel in the integral (\ref{S_ssh}) will have the following form in this basis

\begin{equation}
    \omega \delta_{m,m'} - \hat{H}^{\text{SSH}}_{m,m'} - \hat{V} \delta_{m,m'} \hspace{0.25cm} \rightarrow \hspace{0.25cm} 
    \begin{pmatrix}
        \omega - \varepsilon_{g} & - \Delta \\ 
        - \bar{\Delta} & \omega + \varepsilon_{g}
    \end{pmatrix}
\end{equation}
From where we easily find the eigenvalues $\varepsilon = \pm \sqrt{\varepsilon_{g}^{2} + |\Delta|^2}$. Referring to the statement of the article \cite{BdG} we obtain true energies of corrected edge states

\begin{equation}
\label{Enaive}
    \varepsilon_{1,2} = \sqrt{\varepsilon_{g}^{2} + |\Delta|^2}.
\end{equation}
If we want to consider the energy $\varepsilon_{g}$ inside the superconducting gap, we have an expansion $\varepsilon \approx |\Delta| + \frac{\varepsilon_{g}^2}{2|\Delta|}$.

Now we are ready to compare the results obtained within the phenomenological approach with the previous microscopic analysis. First of all, the naive model (\ref{naive}) does not capture dissipative processes, either due to the quasiparticle leakage inside the superconductor (for energies lying outside the gap) or due to the excitation of the collective modes. Therefore, it cannot be used to estimate the lifetimes of the excitations. Secondly, the naive model exhibits an additional symmetry (\ref{naive_symmetry}) absent in the underlying microscopic model. As a result, the naive model has a degenerate spectrum at $\varepsilon_g = 0$. This unphysical degeneracy can be fixed by introducing the cross-subsystem symmetry-breaking superconducting terms of form
\begin{equation}
    \sum\limits_{n = 0}^{N-1} \left[\Delta^\prime \hat{a}^{\dagger}_{n,\uparrow}\hat{b}^{\dagger}_{n,\downarrow} + \Delta^\prime \hat{b}^{\dagger}_{n,\uparrow}\hat{a}^{\dagger}_{n+1,\downarrow}\right] + h.c.
\end{equation}
which requires introduction of another parameter $\Delta^\prime$. Finally, the leading-order corrections to the quasiparticle spectrum in most cases come from the non-superconducting sectors of $V_{eff}$ (see Eq.(\ref{V_eff})) which leads to a qualitatively different behavior as we change $\varepsilon_g$: in the naive model the corrections are quadratic in $\Delta$ for $\varepsilon \gg \Delta$ while the microscopic model shows that they all are of order $t_0^2$ (Eq.(\ref{edge_state_correction_1})). In order to ensure that the naive model produces similar functional dependencies, one would need to add further terms. It is also worth mentioning that the correction in the naive model always widens the gap, while the microscopically calculated correction may have any sign and tends to be negative at $k = 0$.

\section{Discussion}

In this paper, a microscopic theory of the proximity effect in the contact of a massive s-wave superconductor with a one-dimensional topological insulator described by the Su-Schrieffer-Heeger (SSH) model was constructed using the method of functional integration. The main result is the derivation of an effective action (\ref{S_K}) for the SSH chain, which takes into account induced superconducting correlations through an effective potential $\hat{V}_{m,m'}(\omega)$, nonlocal in both time and space and given by the expression (\ref{V_{int}}). The expression for the action in its general form was obtained regardless of a particular form of the material's quasiparticle spectrum and may serve as a good starting point to describe the proximity effect with a different material, e.g., an unconventional superconductor. Our result enables a consistent analysis of the spectrum of both the bulk and edge states of such a system. The resulting effective action offers several advantages over the widely used approach, which artificially introduces a superconducting order parameter into the Hamiltonian of a topological insulator: it accounts for the potential dissipation, incorporates spacial non-locality which affects the boundary condition for the electron wavefunction and may, therefore, also significantly affect the edge state.

\begin{enumerate}
    \item \textbf{The regime inside the superconducting gap ($|\Delta| > \omega_{0}$)}: As shown in Section 4, in this case, in the absence of interactions the imaginary part of the superconductor's Green's function (\ref{Imw}) vanishes, making the effective operator $\hat{V}_{\omega, k}$ hermitian. This physically implies the absence of dissipation channels for low-energy excitations of the circuit, since the superconductor lacks states into which tunneling could occur. Perturbative calculations yield a correction to the spectrum of bulk excitations (\ref{bulk_excitation}), which removes the band degeneracy. It is important to emphasize that the leading-order correction comes from the normal blocks of the effective potential and scales as $t_0^2$ in tunneling while the anomalous blocks only yield corrections proportional to $t_0^4$. The latter dominate only when the edge state is well localized and at the same time the superconductor exhibits particle-hole symmetry.
    \\
    \item \textbf{The regime outside the superconducting gap ($|\Delta| < \omega_{0}$)}: In this limit, similar to a contact with a normal metal, the effective interaction becomes non-Hermitian. This is manifested by the appearance of a nonzero imaginary part in the corrections to the spectrum (\ref{im}), which corresponds to the finite lifetime of quasiparticles in the chain $\tau \sim 1/(t_{0}^2a^3\nu_{0})$ due to their leakage into the contacting material. Thus, our model adequately describes the transition from the coherent to the dissipative regime with increasing excitation energy, which is fundamentally important for analyzing the transport properties and stability of excitations.
\end{enumerate}

Of particular interest is the influence of the proximity effect on the edge state of a topological insulator. As shown in Section 5, since the energy of the edge state $\varepsilon_{g}$ lies inside the superconductor gap, its localization is preserved. The spacial non-locality of the effective interaction turns out to be insignificant at small transparencies of the barrier. We obtained the shift of the energy of the edge state (\ref{edge_state_correction_1},\ref{edge_state_correction_2}) due to virtual tunneling processes. The resulting expression for the correction qualitatively coincides with the result for a single-level quantum dot associated with a superconductor \cite{Arseev}. The similarity between the two systems can be attributed to the fact that the edge state of the SSH model lies within its band gap and is separated from other states.

We also considered a widely used phenomenological approach in which the order parameter term is added directly to the Hamiltonian of the SSH chain (\ref{naive}). Besides being unable to describe dissipation outside the superconducting gap and spatial nonlocality (which are unimportant for sufficiently low transparencies when the energy of the edge state lies within the superconducting gap), this approach  predicts erroneous dependencies of the energy shifts on the tunneling amplitude. Even if we are not interested in a quantitative picture, the naive approach subtly introduces an additional symmetry (\ref{naive_symmetry}) absent in the microscopic model and leads to an unphysical degeneracy of the energy levels. This degeneracy needs to be treated by adding further symmetry-breaking terms.

Finally, we showed that interactions modify the above picture if the superconductor is quasi-low-dimensional and hosts a soft phase collective mode. In this situation, the tunneling between the materials may excite or absorb such soft modes. At any finite temperature these modes can supply the energy needed for the quasiparticle mode of the chain to escape to the superconductor's quasiparticle band which leads to eventual destabilization of the chain's modes. We also gave an estimate for the lifetime due to this effect, see Eq.(\ref{lifetime}). Overall, while the naive effective model is capable of providing a qualitative picture of how the proximity effect affects the electron states of the SSH chain upon introduction of symmetry-breaking terms, the dissipative effects, the dependence on $\varepsilon_g$ and the functional form of the corrections require a more detailed microscopic calculation.

\section*{Acknowledgments}
The work of A. A. Radkevich was supported by the Basis Foundation under Grant no. 23-1-4-49-1.

\bibliographystyle{unsrturl}
\bibliography{bibliography.bib}

\appendix
\setcounter{equation}{0} 
\renewcommand{\theequation}{А.\arabic{equation}} 

\section{Calculation of the corrections to the bulk energy spectrum}

Here we present the procedure of calculation of the corrections to the energy spectrum of the unperturbed Hamiltonian (\ref{H_k}) due to effective interaction (\ref{V_eff}).

The set of eigenvectors of the Hamiltonian (\ref{H_k})

\begin{equation}
    \{|1\rangle, |2\rangle, |3\rangle, |4\rangle\} = 
    \frac{1}{\sqrt{2}\omega_{0}}
    \begin{pmatrix}
        t _{1} + t_{2}e^{-2ika} & t_{1} + t_{2} e^{-2ika} & 0 & 0 \\
        -\omega_{0} & \omega_{0} & 0 & 0 \\
        0 & 0 & t_{1} + t_{2} e^{-2ika} & t_{1} + t_{2} e^{-2ika} \\
        0 & 0 & -\omega_{0} & \omega_{0}
    \end{pmatrix}
\end{equation}
defined as $\hat{H}^{\text{SSH}}_{k} |n\rangle = \omega_{n}|n\rangle$. In the case where the binding energy is not zero, the unperturbed spectrum has four branches, so in this situation we use non-degenerate perturbation theory. To obtain an answer in the first order in the smallness of the parameter, we need to calculate only the diagonal matrix elements of the effective interaction. Define the matrix element as $V_{ij}(\omega) = \langle i | \hat{V}_{\text{eff}}(\omega)|j\rangle$. So the corrections will have the form

\begin{equation}
    \begin{gathered}
        \omega_{1} = \omega_{0} + \varepsilon_{g} + V_{11}(\omega_{0}+\varepsilon_{g}) \\
        \omega_{2} = -\omega_{0} + \varepsilon_{g} + V_{22}(-\omega_{0}+\varepsilon_{g}) \\
        \omega_{3} = - \omega_{0} - \varepsilon_{g} + V_{33}(-\omega_{0}-\varepsilon_{g}) \\
        \omega_{4} = \omega_{0} - \varepsilon_{g} + V_{44}(\omega_{0}-\varepsilon_{g})
    \end{gathered}
\end{equation}
Calculating the matrix elements we obtain

\begin{equation}
\begin{gathered}
    \omega_1 = \omega_0 + \varepsilon_g + t_0^2 \int\frac{a^2d^2 \bm{k}_{\perp}}{(2\pi)^2}\left[G_{k}^{\uparrow\uparrow}+G^{\uparrow\uparrow}_{k+\frac{\pi}{a}}-\left(G_{k}^{\uparrow\uparrow} -G^{\uparrow\uparrow}_{k+\frac{\pi}{a}}\right)\frac{\cos{(ka)}(t_1+t_2)}{\omega_0}\right]_{\omega = \omega_0 + \varepsilon_g}
\end{gathered}
\end{equation}

\begin{equation}
\begin{gathered}
    \omega_2 = - \omega_0 + \varepsilon_g + t_0^2 \int\frac{a^2d^2 \bm{k}_{\perp}}{(2\pi)^2}\left[G_{k}^{\uparrow\uparrow}+G^{\uparrow\uparrow}_{k+\frac{\pi}{a}}+\left(G_{k}^{\uparrow\uparrow} -G^{\uparrow\uparrow}_{k+\frac{\pi}{a}}\right)\frac{\cos{(ka)}(t_1+t_2)}{\omega_0}\right]_{\omega = - \omega_0 + \varepsilon_g}
\end{gathered}
\end{equation}

\begin{equation}
\begin{gathered}
    \omega_3 = - \omega_0 - \varepsilon_g + t_0^2 \int\frac{a^2d^2 \bm{k}_{\perp}}{(2\pi)^2}\left[G_{k}^{\downarrow\downarrow}+G^{\downarrow\downarrow}_{k+\frac{\pi}{a}}-\left(G_{k}^{\downarrow\downarrow} -G^{\downarrow\downarrow}_{k+\frac{\pi}{a}}\right)\frac{\cos{(ka)}(t_1+t_2)}{\omega_0}\right]_{\omega = - \omega_0 - \varepsilon_g}
\end{gathered}
\end{equation}

\begin{equation}
\begin{gathered}
    \omega_4 =  \omega_0 - \varepsilon_g + t_0^2 \int\frac{a^2d^2 \bm{k}_{\perp}}{(2\pi)^2}\left[G_{k}^{\downarrow\downarrow}+G^{\downarrow\downarrow}_{k+\frac{\pi}{a}}+\left(G_{k}^{\downarrow\downarrow} -G^{\downarrow\downarrow}_{k+\frac{\pi}{a}}\right)\frac{\cos{(ka)}(t_1+t_2)}{\omega_0}\right]_{\omega = \omega_0 - \varepsilon_g}
\end{gathered}
\end{equation}
Substituting the expressions for the Green's function from Eq.(\ref{G}) we obtain

\begin{equation}
\begin{gathered}
    \omega_{1} = \omega_{0} + \varepsilon_{g} + t_0^2 \int\frac{a^2d^2 \bm{k}_{\perp}}{(2\pi)^2}\left[\frac{\omega_{0} + \varepsilon_{g} + \xi_{k}}{(\omega_{0} + \varepsilon_{g} + i 0)^2 - \xi_{k}^2 - |\Delta|^2}+\frac{\omega_{0} + \varepsilon_{g} + \xi_{k+\frac{\pi}{a}}}{(\omega_{0} + \varepsilon_{g} + i 0)^2 - \xi_{k+\frac{\pi}{a}}^2 - |\Delta|^2}\right] - \\ -
    t_0^2 \int\frac{a^2d^2 \bm{k}_{\perp}}{(2\pi)^2} \left[\frac{\omega_{0} + \varepsilon_{g} + \xi_{k}}{(\omega_{0} + \varepsilon_{g} + i 0)^2 - \xi_{k}^2 - |\Delta|^2}-\frac{\omega_{0} + \varepsilon_{g} + \xi_{k+\frac{\pi}{a}}}{(\omega_{0} + \varepsilon_{g} + i 0)^2 - \xi_{k+\frac{\pi}{a}}^2 - |\Delta|^2}\right]\frac{(t_{1}+t_{2})\cos{(ka)}}{\omega_{0}}.
\end{gathered}
\end{equation}

\begin{equation}
\begin{gathered}
    \omega_{2} = - \omega_{0} + \varepsilon_{g} + t_0^2 \int\frac{a^2d^2 \bm{k}_{\perp}}{(2\pi)^2}\left[\frac{-\omega_{0} + \varepsilon_{g} + \xi_{k}}{(-\omega_{0} + \varepsilon_{g} + i 0)^2 - \xi_{k}^2 - |\Delta|^2}+\frac{-\omega_{0} + \varepsilon_{g} + \xi_{k+\frac{\pi}{a}}}{(-\omega_{0} + \varepsilon_{g} + i 0)^2 - \xi_{k+\frac{\pi}{a}}^2 - |\Delta|^2}\right] + \\ +
    t_0^2 \int\frac{a^2d^2 \bm{k}_{\perp}}{(2\pi)^2} \left[\frac{-\omega_{0} + \varepsilon_{g} + \xi_{k}}{(-\omega_{0} + \varepsilon_{g} + i 0)^2 - \xi_{k}^2 - |\Delta|^2}-\frac{-\omega_{0} + \varepsilon_{g} + \xi_{k+\frac{\pi}{a}}}{(-\omega_{0} + \varepsilon_{g} + i 0)^2 - \xi_{k+\frac{\pi}{a}}^2 - |\Delta|^2}\right]\frac{(t_{1}+t_{2})\cos{(ka)}}{\omega_{0}}.
\end{gathered}
\end{equation}

\begin{equation}
\begin{gathered}
    \omega_{3} = - \omega_{0} - \varepsilon_{g} + t_0^2 \int\frac{a^2d^2 \bm{k}_{\perp}}{(2\pi)^2}\left[\frac{-\omega_{0} - \varepsilon_{g} - \xi_{k}}{(-\omega_{0} - \varepsilon_{g} + i 0)^2 - \xi_{k}^2 - |\Delta|^2}+\frac{-\omega_{0} - \varepsilon_{g} - \xi_{k+\frac{\pi}{a}}}{(-\omega_{0} - \varepsilon_{g} + i 0)^2 - \xi_{k+\frac{\pi}{a}}^2 - |\Delta|^2}\right] - \\ -
    t_0^2 \int\frac{a^2d^2 \bm{k}_{\perp}}{(2\pi)^2} \left[\frac{-\omega_{0} - \varepsilon_{g} - \xi_{k}}{(-\omega_{0} - \varepsilon_{g} + i 0)^2 - \xi_{k}^2 - |\Delta|^2}-\frac{-\omega_{0} - \varepsilon_{g} - \xi_{k+\frac{\pi}{a}}}{(-\omega_{0} - \varepsilon_{g} + i 0)^2 - \xi_{k+\frac{\pi}{a}}^2 - |\Delta|^2}\right]\frac{(t_{1}+t_{2})\cos{(ka)}}{\omega_{0}}.
\end{gathered}
\end{equation}

\begin{equation}
\begin{gathered}
    \omega_{4} = \omega_{0} - \varepsilon_{g} + t_0^2 \int\frac{a^2d^2 \bm{k}_{\perp}}{(2\pi)^2}\left[\frac{\omega_{0} - \varepsilon_{g} - \xi_{k}}{(\omega_{0} - \varepsilon_{g} + i 0)^2 - \xi_{k}^2 - |\Delta|^2}+\frac{\omega_{0} - \varepsilon_{g} - \xi_{k+\frac{\pi}{a}}}{(\omega_{0} - \varepsilon_{g} + i 0)^2 - \xi_{k+\frac{\pi}{a}}^2 - |\Delta|^2}\right] + \\ +
    t_0^2 \int\frac{a^2d^2 \bm{k}_{\perp}}{(2\pi)^2} \left[\frac{\omega_{0} - \varepsilon_{g} - \xi_{k}}{(-\omega_{0} + \varepsilon_{g} + i 0)^2 - \xi_{k}^2 - |\Delta|^2}-\frac{\omega_{0} - \varepsilon_{g} - \xi_{k+\frac{\pi}{a}}}{(\omega_{0} - \varepsilon_{g} + i 0)^2 - \xi_{k+\frac{\pi}{a}}^2 - |\Delta|^2}\right]\frac{(t_{1}+t_{2})\cos{(ka)}}{\omega_{0}}.
\end{gathered}
\end{equation}
These expressions show the presence of correspondence between spectrum branches: $\omega_{1} = - \omega_{3}$ and $\omega_{2} = - \omega_{4}$. This is not surprising, since this is a manifestation of the chiral symmetry that our system possesses. The final result for the perturbed spectrum in linear order in $t_{0}^2$ has the form

\begin{equation}
    \omega_{1} = - \omega_{3} = \varepsilon_{g} + \omega_{0} + I(k,\omega_{0}) + I\left(k + \frac{\pi}{a}, \omega_{0}\right)
\end{equation}

\begin{equation}
    \omega_{2} = - \omega_{4} = \varepsilon_{g} - \omega_{0} + I(k,-\omega_{0}) + I\left(k + \frac{\pi}{a}, -\omega_{0}\right),
\end{equation}
where the integral $I(k,\pm\omega_{0})$ is defined as fallows

\begin{equation}
    I(k,\pm\omega_{0}) = t_{0}^2\left[1 \mp \frac{(t_{1}+t_{2})\cos{(ka)}}{\omega_{0}}\right] \int \frac{a^2 d^2 \bm{k}_{\perp}}{(2\pi)^2} \frac{\pm\omega_{0} + \varepsilon_{g} + \xi_{k,\bm{k}_{\perp}}}{(\pm\omega_{0} + \varepsilon_{g}+i0)^2 - \xi_{k,\bm{k}_{\perp}}^2 - |\Delta|^2}. 
\end{equation}
When energy is outside the superconducting gap, the corrections have an imaginary part. For example, for the branch with energy $\omega_{1}$, the expression for the imaginary part has the following form

\begin{equation}
\begin{gathered}
\label{im}
    \text{Im}(\omega_{1}) = - \pi t_{0}^2 \int \frac{a^2 d^2\bm{k}_{\perp}}{(2\pi)^2} (\omega_{0}+\varepsilon_{g}+\xi_{k,\bm{k}_{\perp}})\delta\left((\omega_{0}+\varepsilon_{g})^2 - \xi_{k,\bm{k}_{\perp}}^2 - |\Delta|^2\right)\left[1 - \frac{(t_{1}+t_{2})\cos{(ka)}}{\omega_{0}}\right] - \\
    - \pi t_{0}^2 \int \frac{a^2 d^2\bm{k}_{\perp}}{(2\pi)^2} (\omega_{0}+\varepsilon_{g}+\xi_{k+\frac{\pi}{a},\bm{k}_{\perp}})\delta\left((\omega_{0}+\varepsilon_{g})^2 - \xi_{k+\frac{\pi}{a},\bm{k}_{\perp}}^2 - |\Delta|^2\right)\left[1 + \frac{(t_{1}+t_{2})\cos{(ka)}}{\omega_{0}}\right]
\end{gathered}
\end{equation}
where we used the Sokhotsky theorem. This integral can be evaluated to obtain the following expression for the imaginary part

\begin{equation}
    \text{Im}(\omega) \sim - t_{0}^2 a^3\nu_{0},
\end{equation}
Finally, the lifetime of a quasi-particle can be expressed through the imaginary part using the standard uncertainty relation $\tau \sim - \frac{1}{\text{Im} (\omega)} \sim \frac{1}{t_{0}^2a^3 \nu_{0}}$.

Now we move on to the case with zero binding energy. In this case unperturbed spectrum branches merge into two branches, symmetrical with respect to zero-energy value. Now $\omega_{1} = \omega_{4} = \omega_{0}$ and $\omega_{2} = \omega_{3} = -\omega_{0}$. So for calculation of the correction we should use the degenerate perturbation theory. Then, in the linear order we obtain the following corrections

\begin{equation}
    \omega_{1/4} = \omega_{0} + \frac{V_{11}(\omega_{0}) + V_{44}(\omega_{0})}{2} \pm \frac{\sqrt{(V_{11}(\omega_{0}) - V_{44}(\omega_{0}))^2 + 4|V_{14}(\omega_{0})|^2}}{2}.
\end{equation}

\begin{equation}
    \omega_{2/3} = -\omega_{0} + \frac{V_{22}(-\omega_{0}) + V_{33}(-\omega_{0})}{2} \pm \frac{\sqrt{(V_{2}(-\omega_{0}) - V_{33}(-\omega_{0}))^2 + 4|V_{23}(-\omega_{0})|^2}}{2}.
\end{equation}
where 

\begin{equation}
    V_{14}(\omega) = - V_{23}(\omega) = \frac{i(t_{1}-t_{2})\sin{(ka)}}{\omega_{0}} t_{0}^2 \int \frac{a^2 d^2 \bm{k}_{\perp}}{(2\pi)^2}\left[G_{k}^{\uparrow\downarrow}(\omega)-G^{\uparrow\downarrow}_{k+\frac{\pi}{a}}(\omega)\right].
\end{equation}
It is easy to check that in this case, too, a correspondence arises between the branches of the spectrum: $\omega_{1} = - \omega_{3}$ and $\omega_{2} = - \omega_{4}$, where we use that $V_{33}(-\omega_{0}) = - V_{11}(\omega_{0})$ and $V_{22}(-\omega_{0}) = - V_{44}(\omega_{0})$. As previously, this is connected with the chiral symmetry of our system. We will not write out the final expressions for the corrections with substituted matrix elements due to the large volume of expressions.

\end{document}